\documentclass{JHEP}  
\def\be{\begin{equation}}  
\def\ee{\end{equation}} 
 
\def\bea{\begin{eqnarray}} 
\def\eea{\end{eqnarray}} 
 
\def\ba{\begin{array}} 
\def\ea{\end{array}}

\def\nn{\nonumber} 
\def\Tr{{\rm Tr}}

\def\N4{${\cal N}=4$} 
\def\AdSS5{$AdS_5$} 
\def\AdS5s5{$AdS_5 \times S^5$} 
\def\AdSs5{$AdS_5$} 
\def\AdS5s5{$AdS_5 \times S^5$} 
\def\calP{{\cal P}}

\def\calR{{\cal R}}

\def\calP{{\cal P}}

\def\calA{{\cal A}} 
 
\def\calQ{{\cal Q}}

\def\c1{{\chi^1}}

\def\t2{\tau_2} 
\newbox\SlashedBox 
\def\fs#1{\setbox\SlashedBox=\hbox{#1} 
\hbox to 0pt{\hbox to 1\wd\SlashedBox{\hfil/\hfil}\hss}{#1}} 
\def\hboxtosizeof#1#2{\setbox\SlashedBox=\hbox{#1} 
\hbox to 1\wd\SlashedBox{#2}} 

\def\ms#1{\setbox\SlashedBox=\hbox{$#1$} 
\hbox to 0pt{\hbox to 1\wd\SlashedBox{\hfil/\hfil}\hss}#1} 
 
\def\IZ{\relax\ifmmode\mathchoice 
{\hbox{\cmss Z\kern-.4em Z}}{\hbox{\cmss Z\kern-.4em Z}} 
{\lower.9pt\hbox{\cmsss Z\kern-.4em Z}} {\lower1.2pt\hbox{\cmsss 
Z\kern-.4em Z}}\else{\cmss Z\kern-.4em Z}\fi}

\title{Instantons and  BPS Wilson loops} 
 
\author{Massimo Bianchi\thanks{On leave of absence from 
{\it Dipartimento di Fisica, \ Universit{\`a} di Roma \ 
``Tor Vergata''}, {\it Via della Ricerca  Scientifica 1}, 
{\it 00133 \ Roma, \ ITALY}}, Michael B. Green and Stefano Kovacs \\ 
Department of Applied Mathematics and Theoretical Physics \\ 
CMS, Wilberforce Road, Cambridge CB3 0WA, UK \\ 
E-mail: \email{M.Bianchi@damtp.cam.ac.uk}, 
\email{M.B.Green@damtp.cam.ac.uk}, 
\email{S.Kovacs@damtp.cam.ac.uk}}

\abstract{The 
one-instanton contribution to a circular BPS Wilson loop in 
\N4\ $SU(2)$ Yang--Mills theory is evaluated in semiclassical 
approximation.  This article amplifies part 
of a talk given by MBG at the Strings 2001 conference, Mumbai, India 
(January 5-10, 2001).  The results are preliminary and 
a more complete exposition will be contained 
in a forthcoming paper.}

\preprint{DAMTP-2001-57; ROM2F/2001/27; hep-th/0107028}

\begin{document} 
 
{\it \ldots and having heard that when a man in a forest thinks he is 
going in a straight line, in reality he is going in a circle, I did my 
best to go in a circle, hoping in this way to go in a straight line. 
For I stopped being half-witted and became sly, whenever I took the 
trouble.  And my head was a storehouse of useful knowledge.  And if I 
did not go rigorously in a straight line, with my system of going in a 
circle, at least I did not go in a circle, and that was something.} 
\qquad From `Molloy' by Samuel Beckett

\section{Introduction} 
\label{intro}

The correspondence between Yang--Mills theory and string theory 
embodied in the AdS/CFT conjecture has proved a fertile framework for 
stimulating ideas in quantum gravity as well as in conformal field 
theory.  There is by now much  evidence for the validity of the 
conjecture based on perturbative and non-perturbative calculations.  This 
includes the detailed correspondence between the effects of 
D-instantons  that appear in the derivative expansion (the $\alpha'$ 
expansion) of the 
type IIB superstring theory  and 
Yang--Mills instantons of \N4\ $SU(N)$ supersymmetric Yang-Mills 
theory in the large $N$ limit.  This agreement is particularly 
intriguing since the instanton calculations are semi-classical (weak 
Yang--Mills coupling) while, when the low energy supergravity 
 approximation to string theory is used, 
 the AdS/CFT correspondence should only work at strong coupling. 
 
Tests of the AdS/CFT correspondence involve comparison of observables 
in  string theory with corresponding observables in Yang--Mills 
theory.  Although a great deal of effort has been devoted to 
comparison of amplitudes in type IIB supergravity with correlation 
functions of gauge-invariant composite operators in the Yang--Mills 
theory, the properties of Wilson loops are less well established.  A 
Wilson loop in the large-$N$ gauge theory should be described by a 
minimal surface embedded in $AdS_5$ with a boundary that is a curve on 
the four-dimensional boundary.  The work to be described below 
considers the effect of instantons on a circular  BPS Wilson loop of the \N4\ 
Yang--Mills theory.  This  talk  will only sketch the calculation, 
details of which  will appear shortly \cite{bgk} 
\footnote{The conference talk also reviewed 
results by MBG and SK concerning $1/N$ corrections to instanton 
induced correlation functions and discussed their AdS 
correspondence \cite{gk}.}. 
For our purposes the theory will be considered 
with euclidean time, which means that $AdS_5$ may be considered to be a 
ball $B_5$  with boundary $S^4$.

In a gauge theory with extended supersymmetry 
it is natural to consider a 
special class of Wilson loops that satisfy a BPS condition 
\cite{mald,ry,odg}.  In  \N4\  supersymmetric Yang-Mills theory 
the BPS Wilson loop is a functional of a closed curve, $C$, that is 
defined by \cite{mald,ry,go,odg} 
\be 
\langle W(C)\rangle  = {1\over N} \langle {\rm Tr} {\cal P} \exp 
{\{i\int_{C} (A_{\mu} \dot{x}^{\mu} + [\bar\theta_{A}\dot{x}_{\mu} 
\bar\sigma^{\mu} \lambda^{A} + \theta^{A}\dot{y}_{i} {\widehat
\Gamma}^{i}_{AB}  \lambda^{B} + h.c.] + 
i \varphi_{i}\dot{y}^{i}) ds\} } \rangle \: , 
\label{wildef} 
\ee 
where $\lambda^A$ and $\varphi_i$ are the four Weyl fermion and six 
real scalar fields in the ${\cal N}=4$ supermultiplet and 
${\widehat \Gamma}^{i}_{AB}$ are $SO(6)$ gamma matrices. 
The trace is in the fundamental representation of $SU(N)$ 
and the symbol ${\cal 
P}$ indicates path ordering of the gauge group contractions 
around the loop.  The loop can be parameterized by $x^{\mu} = 
x^{\mu} (s)$ with $s\in [0,1]$ and $x^{\mu} (0) = x^{\mu} (1)$. 
The six coordinates $y^i$ are conjugate to the 
central charges   of the ${\cal N}=4$ superalgebra. 
  In the following we shall be concerned with the 
special class of loops with $\theta^A(s)=0$. In this case if $|\dot{x}| = 
|\dot{y}|$  the expression (\ref{wildef}) satisfies a BPS 
condition, which means that it is invariant under half of the 32 
superconformal symmetries.   More generally, the BPS condition 
allows  $\theta^{A}$ to be freely shifted by any Killing spinor 
$\kappa^{A}$.  This can be seen by exhibiting the $\kappa$ symmetry 
of the Wilson loop \cite{odg}.   The residual supersymmetries are 
defined by the relation between the spinor parameters 
$\kappa^A_\alpha(x)$ and $\bar\kappa_A^{\dot \alpha}(x)$ (which are 
linear functions of $x^\mu$), 
\be 
\dot{x}_{\mu} \sigma^{\mu} \bar\kappa_{A} 
= \dot{y}_{i} {\widehat \Gamma}^{i}_{AB} \kappa^{B} \: . 
\label{kiljoy} 
\ee 
The expression (\ref{wildef}) can be obtained by viewing it as the holonomy 
of an infinitely massive W boson arising from 
the spontaneous symmetry breaking $SU(N+1) \to SU(N) \times U(1)$. 
 
We will further restrict our attention to circular Wilson loops of 
radius $R$ centred at the origin in 
the  $(x^1,x^2)$ plane defined by 
\be 
(x^1)^2 + (x^2)^2 = R^{2},\qquad x^3 (s) = 0 , \qquad x^0 (s) = 0, \qquad 
\dot{y}^{i}(s) = |\dot{x}| n^{i} \: , 
\label{circloop} 
\ee 
where $n^{i}$ is a unit vector in the six `internal' $SU(4)$ 
directions. The superconformal invariance of the theory 
guarantees that the Wilson loop,  $\langle W(R)\rangle$, is a constant that is 
independent of $R$.  Some perturbative contributions to Wilson loops of this 
kind (namely, the `rainbow' diagrams) 
have been calculated to all orders in the coupling constant 
\cite{esz} and argued to be exact, at least in the large $N$ limit. 
A clever anomaly argument in \cite{dg} uses  the fact that a 
straight line Wilson loop has no perturbative 
contributions to determine the expression for a circular Wilson loop. 
The conformal transformation that maps 
the line to a circle is anomalous and generates a nontrivial 
expression for $\langle W(R)\rangle$.  Although this has been conjectured to 
be correct not only to all orders in perturbation theory but also to 
 all orders in the 
$1/N$ expansion it has been argued that the $1/N$ corrections are not 
universal \cite{ad}.  The result is consistent 
with expectations based on the AdS/CFT correspondence.  Our aim is 
to compute the 
one-instanton contribution to the circular Wilson loop.  We 
will work to lowest order in the 
Yang--Mills coupling constant, $g_{_{YM}}$ (the semiclassical 
approximation)  and  take the 
gauge group to be $SU(2)$, although it is very straightforward to 
generalize this to $SU(N)$ for arbitrary $N$.  The very notion of a 
`straight line Wilson loop' with sensible boundary conditions 
is problematic in the context of the instanton calculation so we will 
consider the circular case directly. 
 
A single instanton in the \N4\ $SU(2)$ Yang--Mills theory has eight bosonic 
moduli and sixteen fermionic moduli.  The bosonic moduli correspond to 
broken translation, dilation and gauge orientation symmetries. 
The fermionic moduli are in 
one-to-one correspondence with the broken supersymmetries and 
conformal supersymmetries.  In the following we will evaluate 
 the integral over these supermoduli in the background of 
a Wilson loop. Ignoring the three gauge moduli, which 
are irrelevant since the Wilson loop is 
gauge invariant, the bosonic collective 
coordinates $( x_{0}^{\mu}, \rho_{0})$ 
parametrize a copy of $AdS_5$. 
The fermionic collective coordinates 
$(\eta^{A}, \bar\xi^{A})$ can be packaged into the 
chiral spinor, 
\be 
\zeta^{A}(x) = \eta^{A} + x_{\mu} 
\sigma^{\mu} \bar\xi^{A} \: , 
\label{zetdef} 
\ee 
which (up to rescalings) 
 is also a Killing spinor of $AdS_5$, which indicates a 
holographic connection between the Yang--Mills instanton and the 
D-instanton of the IIB string theory in $AdS_5\times S^5$ 
\cite{bg,bgkr}. 
 
The value of the Wilson loop in this background 
is given, to leading order in the 
coupling constant, by substituting the classical instanton solution 
into (\ref{wildef}) and integrating over the supermoduli with the 
appropriate invariant measure.  The `classical' solution must include 
not only the standard instanton solution but also 
the dependence of the fields on fermionic moduli. 
In the instanton background the fields $\varphi^i$ and $A_\mu$ 
have zero modes that depend on the fermionic moduli which are 
induced by the presence of Yukawa couplings in the usual manner. 
The multiplet of these zero modes can be generated from the classical 
instanton profile of the vector potential by successive application of 
the supersymmetries that are broken by the instanton.  Thus, two 
transformations generate 
\be 
\widehat \varphi^{ia} = {1\over 2} F^{a\, -}_{\mu\nu} 
{\widehat \Gamma}^{i}_{AB} \zeta^{A} \sigma^{\mu\nu} \zeta^{B} \: , 
\label{vardef} 
\ee 
where the  hat  indicates the value of a field containing fermionic collective 
coordinates induced by the instanton background 
and $F^{a\, -}_{\mu\nu}$ is the anti self-dual field 
strength of the classical instanton solution.\footnote{We may ignore 
the change in the classical field equations due to the presence of the 
Wilson loop source since such changes enter at higher order in $g_{_{YM}}$.} 
Performing another two supersymmetry transformations gives 
\be 
\widehat A_{\mu} = {1\over 4!} \varepsilon_{ABCD}\zeta^{A} 
\sigma_{\mu\nu} \zeta^{B} D^{\nu} ( F^{a\, -}_{\lambda\kappa} 
\zeta^{C} \sigma^{\lambda\kappa} \zeta^{D}) \: , 
\label{ahdef} 
\ee 
which is a potential that corresponds to a self-dual field 
strength, $\widehat F^{a +}_{\mu\nu}$.  The calculation of the Wilson 
loop to leading order in $g_{_{YM}}$ involves substituting 
the expressions (\ref{vardef}) and (\ref{ahdef}) into 
(\ref{wildef}).  The sixteen fermionic integrals are 
then saturated  by expanding the exponential to extract the terms 
with  sixteen powers of $\zeta$.  This involves dealing with the 
complicated combinatorics of terms with $n$ $\widehat\varphi$'s and 
$m$ $\widehat A$'s with $2 n + 4 m =16$ that arise from expanding the 
exponent in (\ref{wildef}).

\section{Symmetries of  the super Wilson Loop}

\subsection{The bosonic model} 
 
Even in the absence of fermionic moduli it is difficult to 
evaluate the classical instanton contribution to the Wilson loop 
by direct integration over the bosonic moduli. However, it is possible to 
proceed by making use of the symmetries of the problem.  It is 
straightforward to see that the presence of the circular loop 
partially breaks the conformal $SO(4,2)$ symmetry 
to a residual unbroken $SO(2,2)$ subgroup.  The instanton calculation 
will require continuation to euclidean signature in which case the residual 
symmetry is an $SO(3,1)$ subgroup of $SO(5,1)$. 
 
The fact that the instanton is invariant in form under these 
transformations, up to irrelevant gauge transformations, allows 
us to make use of this symmetry in order to map an arbitrary 
instanton to one that is located at the centre of the loop.  The 
expression for the integrand of the loop then becomes abelian and 
it is easy to evaluate the integral explicitly. Of course, such a 
classical calculation is not directly relevant to pure 
Yang--Mills theory since in that case the classical conformal 
symmetry is broken by quantum fluctuations that require 
renormalization.   In this toy example the 
integral diverges since there is no suppression of instantons 
with arbitrarily small scale size or ones that are located 
arbitrarily far from the loop. Nevertheless, it proves useful to 
study this example since it will later be generalized to the case of the 
superconformal \N4\ theory.

The  infinitesimal conformal transformations, 
\be 
\delta x^{\mu} = a^{\mu} + \omega^{\mu\nu} x_{\nu} + \lambda 
x^{\mu} + x^{2} b^{\mu} - 2 b\cdot x x^{\mu} \: , 
\label{confdef} 
\ee 
are generated by $( P_{\mu}, J_{\mu\nu}, D, K_{\mu})$, 
 where $(a^\mu, \omega^{\mu\nu}, \lambda, b^\mu)$ are constant 
parameters. 
In order to streamline the discussion  of the conformal properties 
it proves very convenient to make use of Dirac's formalism 
\cite{dirac} for representing the conformal group by extending 
four-dimensional Minkowski space-time  to six dimensions with 
signature $(4,2)$ and  coordinates $X^M$ ($M=0,\dots,5$), where 
$X^0$ and $X^4$ are time-like. In this way, $SO(4,2)$ can be 
represented linearly by rotations and boosts on $X^M$.  In order 
to find a representation of $SO(4,2)$ in five dimensions the 
flat  six-dimensional coordinates are taken to satisfy the 
invariant constraint, 
\be 
X^2\equiv \eta_{MN} X^M X^N \equiv 
\eta_{\mu\nu} X^\mu X^\nu + (X^4)^2  - (X^5)^2  = C^2 \: , 
\label{sixcon} 
\ee 
where $C$ is an arbitrary constant scale that drops out of the 
conformally invariant theory. For now we are using the 
six-dimensional metric $\eta_{MN} = {\rm diag}(+---+-)$ and the 
four-dimensional metric $\eta_{\mu\nu} = {\rm diag}(+---)$, 
although later we will consider the Wick rotation of $X^0$ that 
is relevant for the instanton calculation. The constraint is 
solved in terms of five-dimensional coordinates that define 
$AdS_5$ with scale $C$. It will prove 
notationally convenient to choose the arbitrary scale to be the radius 
of the Wilson loop, $C=R$, from now on.  A conventional parameterization of 
$AdS_5$ in terms of $x^\mu$ and $\rho$ is obtained by the 
identifications 
\be 
X^\mu = R\, {x^\mu \over \rho},\qquad X^4 = {1\over 2} \left( \rho + 
{R^2 - x^2 \over \rho}\right), \qquad 
X^5 = {1\over 2} \left( \rho - {R^2 + x^2 \over \rho}\right) \: , 
\label{solvec} 
\ee 
which represents an $AdS_5$ hypersurface in ${\bf R}^6$.  
Lorentz 
transformations of the six-dimensional coordinates generated by the 
generalized angular momenta, 
\be 
L_{MN} = X_M 
\partial_N - X_N \partial_M \: , 
\label{sixtrans} 
\ee 
are isomorphic to $SO(4,2)$ transformations acting on $AdS_5$, with the 
identifications of the fifteen generators, 
\be 
J_{\mu\nu} = L_{\mu\nu} 
\, , \qquad D = L_{45} \, , \qquad P_\mu = L_{4\mu} + L_{5\mu} \, , 
\qquad K_\mu = L_{4\mu} - L_{5\mu} \: . 
\ee 
Furthermore the trivial `flat' six-dimensional integration measure is 
equivalent, after the constraint, to the $AdS_5$ measure, 
\be 
R^{-4} \,\int \delta(X^2 -R^2) \,  d^6X = \int  {d^4 x 
d\rho\over \rho^5} \: . 
\label{measeq} 
\ee 
It is easy to verify that the boundary of $AdS_5$ is 
mapped into itself under the $SO(4,2)$ transformations and that the standard 
four-dimensional action of $SO(4,2)$ results from the limit in which 
the radial variable goes to the boundary, $\rho \to 0$ 
\cite{dirac}. 
 
Whereas a generic Wilson loop breaks the $SO(4,2)$ conformal symmetry 
completely, we will make use of the fact that a circular 
Wilson loop leaves an 
$SO(2,2)$ subgroup unbroken. 
This follows from the fact that a loop in the 
$(x^1,x^2)$  plane on the 
boundary of $AdS_5$ at $\rho=0$, $|x_l| = R$, $x_t=0$ is mapped into 
itself under $SO(2,2)$ transformations, 
where $x_l=(x_1,x_2)$ and 
$x_t=(x_0,x_3)$ are longitudinal and transverse vectors.  These $SO(2,2)$ 
transformations are described in the six-dimensional formalism as those that leave 
invariant the quadratic form, 
\be 
U(x,\rho)  = (X^4)^2 - (X^l)^2\equiv X_L^2 
\: , 
\label{bosform} 
\ee 
where $X^L = (X^4, X^l)$.   After 
imposing the constraint (\ref{sixcon}) the quantity 
\be 
V(x,\rho) \equiv X_T^2 = U - R^2 
\label{vdef} 
\ee 
is also invariant, where $X^T = (X^0,X^3,X^5)$. 
 For fixed $U$ (\ref{bosform}) defines a four-dimensional 
hyperbolic surface in 
$AdS_5$ once the constraint (\ref{sixcon}) is imposed.  The 
four-dimensional surfaces of constant $U$ foliate 
$AdS_5$ in such a manner that they all meet on the circle at the 
boundary, $\rho=0$, $x_t=0$, $|x_l|=R$. 
The minimal value of $U = R^2$ is achieved on  a degenerate 
surface that is two-dimensional. This surface is, in fact, the 
surface of minimal area in $AdS_5$ that bounds the loop of radius 
$R$ on the boundary that was described in \cite{mald,odg}. 
It follows from (\ref{bosform}) that 
a general point in $AdS_5$ has the same value of $U$ as a point 
at the centre of the loop with $\tilde x^\mu =0$ and  $\tilde 
\rho(x,\rho)$  determined by 
\be 
U(\tilde x =0,\tilde \rho) = 
{(R^2 + \tilde\rho^2)^2 \over 4\tilde\rho^2} 
= {(R^2 - x_l^2 - x_t^2 + \rho^2)^2 \over 4\rho^2} + 
{R^2 x_l^2\over \rho^2} = U(x,\rho) \: . 
\label{invcent} 
\ee 
We will later interpret $AdS_5$ as the moduli space of an $SU(2)$ instanton. 
 
\subsection{The ${\cal N}=4$ supersymmetric theory} 
 
We are now interested in extending this discussion to the \N4\ 
supersymmetric theory which is invariant under 
the supergroup $SU(2,2|4)$.  The bosonic part of this group is 
$SO(4,2) \times SU(4)$.  The fifteen transformations of the 
$SU(4) \approx SO(6)$ R-symmetry  group (with parameters 
$\omega^{ij}$) have the form, $\delta y^{i} = \omega^{ij} y_{j}$. 
In addition, there are   four Poincar\'e supersymmetries with 
generators $( Q_{A}^{\alpha}, \bar{Q}^{A}_{\dot\alpha})$, 
and four superconformal symmetries with generators 
$(S^{A\alpha}$,  $\bar{S}_{A\dot\alpha})$. 
 
We can extend Dirac's six-dimensional representation of 
$SO(4,2)$ to $SU(2,2|4)$ by including an appropriate set of 
fermionic coordinates.  There is a wide variety of possible 
choices for such coordinates but we 
will choose a `chiral' representation that is 
particularly well adapted for the 
instanton calculation and requires euclidean signature.  
Recall that we wish to integrate over the 
sixteen components of the broken supersymmetries and conformal 
supersymmetries, $\eta^A_\alpha$ and $\bar \xi^{\dot\alpha A}$. 
Therefore, in addition to the six coordinates $X^M$, we now include a 
quartet of Grassmann spinors, $\Theta^A_a$, where $a$ is a 
four-component spinor index.  This will be interpreted as a 
chiral spinor in the six-dimensional description. The 
corresponding ${\cal N}=4$ instanton superspace will be chosen to 
be the supercoset 
\be 
{SU(2,2|4) \over Span \{ SO(4,1)\times SO(5); 
\bar{Q}^A_{\dot\alpha}, S^B_{\alpha} \} } \: , 
\label{supercos} 
\ee 
where the generators in $SO(6)/SO(5)$ will play no role in the 
following and the remaining coordinates are simply the supermoduli 
associated with the instanton. 
 
Using a standard super-coset construction \cite{supcos} 
it is straightforward to 
exhibit the action of $SU(2,2|4)$ on these coordinates \cite{gk}. These 
transformations may be written compactly in a six-dimensional 
notation by the identification 
\be 
\Theta_a^A = ( \eta^A_\alpha + 
x\cdot\sigma_{\alpha\dot\alpha}\bar\xi^{\dot\alpha A} \: , 
\bar\xi^{A}_{\dot\alpha})\: , 
\label{thetdeff} 
\ee 
where $a=(\alpha,\dot\alpha)$ is a spinor index of $SO(4,2)$ (or 
$SO(5,1)$ in the euclidean theory) and takes values  from 1 to 
4.  Our notation anticipates the fact that we will identify the 
sixteen fermionic variables with the sixteen collective 
coordinates of the instanton that are associated with the broken 
superconformal symmetries. The 32 supercharges can be packaged into 
six-dimensional chiral  spinors $\calQ^a_A = (Q_A^\alpha, 
\bar{S}_{A}^{\dot\alpha})$ and 
$\bar\calQ^A_a = (\bar{Q}^A_{\dot \alpha}, S^A_{\alpha})$ and
represented in terms of the supercoordinates by 
\be 
\calQ^a_A = {\partial \over \partial \Theta^A_a},\qquad 
\bar{\calQ}^A_a =  \Theta^A_b \Theta^B_a{\partial \over \partial 
\Theta^A_a} + {1\over 4}\Gamma^{MN}_a{}^b\Theta^A_b L_{MN}\: , 
\label{qdefs} 
\ee 
which satisfy the $SU(2,2|4)$ superalgebra 
\be 
\{ \calQ^a_A, \bar{\calQ}^B_b\} = {1\over 4}\delta^B_A \Gamma^{MN}_b{}^a 
J_{MN} + {1\over 4}\delta^a_b \widehat{\Gamma}^{ij}_A{}^B T_{ij}\: , 
\label{ressdef} 
\ee 
where $J_{MN} = L_{MN} + S_{MN}$ are the generators of $SO(4,2)$ and 
$T_{ij}$ are the generators of $SO(6)$. More explicitly, 
\be 
L_{MN} = X_M \partial_N - X_N \partial_M\: , \qquad 
S_{MN} =  \Theta^A_a \Gamma_{MN}^a{}_b 
{\partial \over \partial \Theta^A_b}\: ,\qquad 
T_{ij} = \Theta^A_a \widehat{\Gamma}_{ijA}{}^B 
{\partial \over \partial \Theta^B_a} \: . 
\label{angsed} 
\ee 
 The  superconformal transformations of the coordinates generated by 
these charges are 
\bea 
\delta \Theta^A_a = \epsilon^A_a\: ,  &&\qquad \bar\delta\Theta^A_a = 
\Theta^A_b \Theta^B_a \bar\epsilon^b_B\: , 
\nn \\ 
\delta X^M = 0, &&\qquad \bar\delta X^M = {1\over 2} 
\bar\epsilon_A \Gamma^{MN}\Theta^A X_M \: . 
\label{supcons} 
\eea

The presence of the Wilson loop breaks this symmetry to a 
 subgroup of  $SU(2,2|4)$ that leaves the 
loop invariant up to reparametrizations.  Since we have fixed a 
direction ($n^{i}$) in the internal space,  the loop is only invariant 
under an $SO(5) \approx Sp(4)$ subgroup of the $SU(4) \approx 
SO(6)$ R-symmetry group. It will prove convenient to define the $Sp(4)$ 
singlet $\Omega_{AB} = n_{i} {\widehat \Gamma}^{i}_{AB}$. 
We now want to find the fermionic part of the superconformal group 
that leaves the loop invariant.  Since the  bosonic symmetry that 
preserves the loop, $SO(2,2)\times Sp(4)$,  is the bosonic part of the 
supergroup $OSp(2,2|4)$ (which is a subgroup of $SU(2,2|4)$)  this is 
a natural candidate for the invariance group of the loop.  To verify 
that this is indeed the case we need to identify the  Killing 
spinors  that satisfy the relation (\ref{kiljoy}). 
It turns out that these correspond to symmetries generated by the 
linear combinations 
\be 
G_{A} = \sigma^{12} Q_{A} + {1\over R}\Omega_{AB} S^{B} 
\label{gdeff} 
\ee 
and its conjugate, $\bar{G}^{A}$.   The index $A$ is raised and 
lowered by the symplectic metric $\Omega_{AB}$. In a compact 
notation the surviving supersymmetry generators $G$ and $\bar{G}$ 
can be packaged into $G_A^a$, where $a$ is an index in the ${\bf 
(2,2)}$ of $SO(2,2)$ and the supersymmetry algebra becomes 
\be 
\{ G^a_A, G^b_B \} = \Omega_{AB} J^{ab} + T_{(AB)} H^{ab} \: , 
\label{compsus} 
\ee 
where the six generators of $SO(2,2)$, 
$( \Pi^+_l, J_{xy},\Pi^-_t, J_{zt})$ have been assembled into 
$J^{ab}$ and $H^{ab}$ is the symmetric $SO(2,2)$ invariant metric. 
The remaining  commutation relations of the $OSp(2,2|4)$ algebra are 
\bea 
{} [J_{ab}, J_{cd}] &=& H_{bc} J_{ad} + {\rm perms}\: , \qquad 
{} [T_{AB}, T_{CD}] = \Omega_{BC} T_{AD} + {\rm perms}\: , \nn \\ 
{} [T_{AB}, J_{ab}] &=& 0\: , \quad\qquad 
\qquad\qquad\qquad {} [J_{ab}, G_{Ac}] = H_{bc} 
G_{Aa} - H_{ac} G_{Ab} \: ,
\nn \\ 
{} [T_{AB}, G_{Ca}] &= &\Omega_{BC} G_{Ac} +  \Omega_{AC} G_{Bc} \: . 
\label{smallalg} 
\eea 
 
\section{One-instanton contribution to the Wilson loop} 
\label{pureYM} 
 
We are now in a position to consider the integral over the 
instanton supermoduli that enters into  the expression for the Wilson 
loop. We have been unable to find any analytic treatment of 
instanton contributions to Wilson loops in the 
literature, even  for the purely bosonic theory, 
although there have been many approximate and 
numerical treatments in the context of (lattice) QCD. 
 
\subsection{The bosonic model} 
 
We will again begin by  first considering the 
 toy conformally invariant model in which 
there are no fermions and  only the bosonic moduli arise. 
The expression for the 
classical instanton contribution to the standard Wilson loop of 
pure $SU(2)$ Yang--Mills theory (ignoring overall constants) is given by an 
 integral over the instanton moduli, 
\bea 
\langle W(R)\rangle_{bos}  &=&  \int d^6 X_0 \, 
\delta( X_0\cdot X_0 - R^2)\, \Tr \calP\, e^{i\int_C A\cdot \dot x ds} \nn \\ 
&=&  \int d^6 X_0 \, \delta(X_0\cdot X_0- R^2)\, 
\Tr \calP\, \exp\left(i\int_C {\eta^a_{\mu\nu} (x^\mu - x^\mu_0) 
\sigma_a \over \rho_0^2 + |x - x_0|^2}\, \dot x^\nu\, ds\right) 
\: ,
\label{bossix} 
\eea
where $\eta^a_{\mu\nu}$ is the standard 'tHooft eta symbol. 
 Since the form of the instanton solution is 
invariant under $SO(4,2)$ transformations  a particular  one-instanton 
configuration can be transformed into an 
equivalent one by acting with an element of $SO(2,2)$, which maps 
the loop onto itself.  In this way an arbitrary instanton with moduli 
 $(x_0,\rho_0)$ can be mapped 
into  the special configuration in which it is at the centre of the 
loop with moduli $(\tilde x_0=0, \tilde 
\rho_0(x_0,\rho_0))$. 
Then the path ordered exponential becomes simple since $\sigma_a 
\eta^a_{\mu\nu} \, x^\mu \, \dot x^\nu ds = R^2 \sigma^3 d\phi$, 
where $0\le \phi\le 2\pi$ is the angle around the loop, which has 
been taken to lie in the $(x^1,x^2)$ plane.  In this way the exponent is 
abelianized and the integration over $\phi$ followed by the trace 
simply leads to 
\be 
\langle W(R)\rangle_{bos} = 
 \int d^6X\, \delta(X^2 - R^2)\, W_B(|X_T|)\: , 
\label{cenres} 
\ee 
(dropping the subscript $0$ from the collective coordinates, $X^M$) 
where the bosonic Wilson loop density is a scalar function of the $SO(2,2)$ 
invariant which can be written as 
\be 
W_B(|X_T|)   =  \cos\left({\pi |X_T| \over \sqrt{R^2 + X_T^2}}\right) 
\label{specip} 
\ee 
after using the constraint $X_L^2 = X_T^2 + R^2$. 
These expressions are 
to be interpreted after a Wick rotation of $X^0$. 
 
 The specific $SO(2,2)$ group element that generates this transformation 
is of the form $\exp(a^l \Pi^+_l  + a^t \Pi^-_t )$ where the 
parameters $a^l(x_0^\mu,\rho_0)$ and $a^t(x_0^\mu,\rho_0)$ are 
functions of the collective coordinates that can be 
determined by elementary group theory. The integral 
(\ref{cenres}) is highly divergent,  which was anticipated in 
this toy model. However, the expectation is that the 
supersymmetry of the \N4\ theory will lead to compensating 
cancellations.

\subsection{Superinvariants} 
 
We saw earlier in 
(\ref{bosform}) how to define quadratic invariants, $U\equiv X_L^2$ 
and $V\equiv X_T^2 = U - R^2$, 
for the group $SO(2,2)$.  Clearly any function of $V$, such as the 
bosonic Wilson loop density $W_B(|X_T|)$ (\ref{specip}) is also 
invariant.  We can  extend this to an invariant 
of $OSp(2,2|4)$ by solving the equation 
\be 
(\epsilon^A_a \calQ^a_A + \bar\epsilon_A^a \bar{\calQ}^A_a) W(X,\Theta) = 0 
\label{ospferm} 
\ee 
with the  condition $W(X,0) = W_B(|X_T|)$. 
The restriction to $OSp(2,2|4)$ is built into the condition 
\be 
\bar\epsilon_A^a = \Omega_{AB} H^{ab} \epsilon_b^B. 
\label{bpsag} 
\ee 
$Sp(4)$ and $SO(2,2)$ indices are raised and lowered by 
the symplectic metric $\Omega_{AB}= n_i \widehat{\Gamma}^i_{AB}$ and 
the symmetric tensor $H^{ab}=\Gamma_{412}^{ab}$, respectively. 
Equation (\ref{ospferm}) imposes the 
condition that the loop is BPS so it is annihilated by the 
fermionic generators  of $OSp(2,2|4)$.  It can be written as 
\be 
{\partial W(X,\Theta) \over \partial \Theta^A_a} + 
\Omega_{AB}H^{ab} \left[ 
 \Theta^B_c \Theta^C_b{\partial W(X,\Theta) \over \partial 
\Theta^C_c} + {1\over 4}\Gamma^{MN}_b{}^c\Theta^B_c L_{MN}W(X,\Theta) 
\right] = 0 \: , 
\label{eqwils} 
\ee 
which can be solved in a standard fashion, giving 
\be 
W(X,\Theta) = \calP \exp\left(-\int_0^1 {du\over 4u} 
f^{MN}(u\Theta)L_{MN} \right) W_B(|X_T|) \: , 
\label{finthet} 
\ee 
where $\calP$ indicates that the exponential is to be path ordered. 
The matrix function $f^{MN}(u\Theta)$ is defined by 
\be 
f^{MN}(\Theta) = \bar\Theta_A \Gamma^{MN} \Theta^A + 
 \bar\Theta_A \Theta^B \bar\Theta_B \Gamma^{MN} \Theta^A 
\label{fdeff} 
\ee 
where as in (\ref{bpsag}) $\bar\Theta_A^a =\Omega_{AB}H^{ab}\Theta^B_b$. 
Two important features of $f^{MN}$ are worth stressing. 
Firstly it is at most quartic in $\Theta$, all higher order terms being 
zero after (anti)symmetrization of the indices implied by the 
structure of the contractions. Secondly it only has 
non-zero entries for the nine elements of the 
coset\footnote{In euclidean signature 
$SO(5,1)/SO(3,1)$} 
$SO(4,2)/SO(2,2)$. In other words the only 
$L_{MN}$'s that enter (\ref{finthet}) are those of the form $L_{ij}$ 
where the index $i=1,2,3$ labels the directions in $X_L$ and $j=1,2,3$ 
labels the directions in $X_T$. 
 
It is notable that the dependence on $\Theta$ in (\ref{finthet}) 
enters via the 
exponential prefactor that acts simply as a rotation on the bosonic 
coordinates, $X^M$. This means that the 
invariant $W(X, \Theta)$ 
is simply obtained  by transforming the coordinates $X^M$ in 
$W_B(|X_T|)$ by  a 
$\Theta$-dependent rotation, 
\be 
X^M \rightarrow \tilde{X}^M = \calR^M{}_N(\Theta) X^N\: , 
\label{varch} 
\ee 
where the rotation matrix, $\calR(\Theta)$, is given by the 
six-dimensional (fundamental) representation of the path-ordered exponential in 
(\ref{finthet}). 
In fact, a naive argument at this point would say that all of the 
dependence on $\Theta$ can be eliminated by simply changing 
integration variables from $X^M$ to $\tilde X^M$.  Since the jacobian 
for this change of variables is trivial,  the resulting Wilson loop 
would clearly vanish after the Grassmann integrations.  However, this 
ignores a crucial subtlety -- the original bosonic integration is 
divergent and has to be regulated near the boundary of $AdS_5$. 
In the presence of a regulator, no matter how it is defined, the 
change of variables $X\to \tilde X$ introduces dependence on $\Theta$ 
into the boundary conditions so the Grassmann integral no longer 
vanishes automatically.

\subsection{The ${\cal N}=4$ supersymmetric Yang--Mills theory} 
 
The expression for the Wilson loop  in the \N4\ $SU(2)$ theory is given 
by the integral (up to an overall constant) 
\be 
\langle W(R) \rangle = 
 \int d^6 X\, \delta(X_L^2 - X_T^2 -R^2)\,\int 
d^{16}\Theta\, W(X,\Theta)\: , 
\label{loopint} 
\ee 
with $W(X,\Theta)$ given by (\ref{finthet}). 
The integral over the supermoduli can be evaluated by expanding 
the exponential in a power series in $\Theta$ in order 
to determine the $\Theta^{16}$ term.   Defining 
\be 
\Phi = \bar\Theta_A \Gamma^{ij}\Theta^A L_{ij} \: , \qquad  \calA = 
\bar\Theta_A \Theta^B \bar\Theta_B \Gamma^{ij}\Theta^A L_{ij}\: , 
\ee 
the $\Theta^{16}$ terms in the expansion of the exponent in 
(\ref{finthet})  can be written schematically as 
\be 
{1\over 8!} \Phi^8 + {1\over 6!2} \Phi^6 \calA + {1\over 4!2!4} \Phi^4 
\calA^2 + {1\over 2!3!8} \Phi^2 \calA^3 + {1\over 4!16} \calA^4\: . 
\label{expans} 
\ee 
The precise coefficients and contractions can be evaluated taking 
into account the fact that the operators $\calA$ and $\Phi$ do not 
commute.  The terms in this expansion are in one-to-one correspondence 
with the terms in the expansion of the original Wilson loop 
(\ref{wildef}) in powers of $\widehat\varphi$ and $\widehat A$ (defined in 
(\ref{vardef}) and (\ref{ahdef}), respectively). 
 
However, it is considerably more laborious 
to evaluate the action of these terms on the function $W_B(|X_T|)$ that 
enters into (\ref{finthet}).  Part of the problem is the evaluation of the 
 Grassmann integration over the sixteen 
fermionic collective coordinates.  The method we have used is to 
decompose the integration into 
product of integrals over two eight-component $SO(8)$ spinors defined 
by 
\be \hat{\theta} = (\Theta^1, \Theta^2)\: , 
\qquad \check{\theta} = (\Theta^3, \Theta^4)\: . 
\label{thetchec} 
\ee 
Then use can be made of the standard expression 
\be 
\int d^8 \theta \theta \gamma^{m_1 n_1} \theta .. \theta 
\gamma^{m_4 n_4} \theta = t_8^{m_1 n_1 .. m_4 n_4} \: ,
\ee 
where $m_i,n_i = 1,...8$ and $t_8$ is a well-known  $SO(8)$ invariant 
tensor \cite{gsa}. In this way, after many algebraic manipulations, 
the Grassmann  integrations in (\ref{loopint}) can be performed and 
result in tensors that contract into products of even powers of the angular momentum 
generators, $L_{ij}$.  It is easy to see that terms 
with an odd number of $L$'s vanish due to the (angular) bosonic integrations 
and therefore can be ignored. 
This however leaves a huge number of terms, 
 each of which consists of a tensor with eight, twelve or sixteen indices 
 contracting into a product of four, six or eight $L$'s which act on 
$W_B(|X_T|)$. 
 In order to evaluate these expressions we have made extensive use 
 of the algebraic software package REDUCE.  The result has the form, 
\bea 
F(|X_T|) &\equiv & 
\int d^{16}\Theta \left[{1\over 8!} \Phi^8 + {1\over 4!2!4} \Phi^4 
\calA^2 + {1\over 4!16} \calA^4\right] W_B(|X_T|)\nn\\ 
& =& \sum_{n=1}^{8} \sum_{k=0}^{\left[{n+2\over 2}\right]} 
C_{n+2-2k}^{(n)}\, |X_T|^{n-2k} R^{2k} \, 
{\partial^n W_B(|X_T|)\over \partial |X_T|^n} \; . 
\label{ewthe} 
\eea 
 
Although we have not yet completed the evaluation of the coefficients 
$C_m^{(n)}$ we can consider general features that arise after 
substituting (\ref{ewthe}) into the expression (\ref{finthet}) for 
the Wilson loop.  The bosonic integral superficially seems 
 to be divergent, at least 
term by term in the expansion (\ref{expans}).  Such divergences arise 
at the boundary of moduli space.  However, we have also 
seen that the fermionic integrations naively lead to a vanishing 
result if the subtleties of the boundary of moduli space are ignored. 
We therefore regulate the divergences by cutting off the 
integration near the boundary.  This is achieved by requiring that 
$X_4 < \Lambda$ so that the bosonic integral becomes 
\bea 
\langle W (R) \rangle_\Lambda &=& \int^\Lambda dX_4 \int 
 d^2X_l  d^3X_T \, \delta(X_4^2 - X_l^2 - 
X_T^2 - R^2) \, F(|X_T|) 
\nn \\ 
& =& 
4\pi^2 \int_0^{\sqrt{\Lambda^2 - R^2}} d|X_T| \, |X_T|^2\,  (\Lambda - \sqrt{X_T^2 + 
R^2})\,  F(|X_T|)\: . 
\label{reswilsb} 
\eea 
As before, this expression is to be evaluated with the choice of 
euclidean signature for $X^\mu$ so that $X_T^2>0$. 
The integration is now performed by substituting the expression 
for $F(|X_T|)$ from (\ref{ewthe}) and using the explicit expression for 
$W(|X_T|)$ in (\ref{specip}).   The integral 
in (\ref{reswilsb})   is 
at most linearly divergent rather than having the quartic 
divergence implied by naive dimensional analysis as in the bosonic model. 
 
The apparent 
linear divergence has a coefficient that should vanish although the 
explicit expression for $F(|X_T|)$ does not make this at all 
apparent. The absence 
of this linear divergence requires 
\be 
\sum_{n=1}^{8} (-)^{n} (n+2)! C_{n+2}^{(n)} = 0\:,\qquad 
\sum_{n=1}^{8} (-)^{n} n! C_{n}^{(n)} = 0\: . 
\label{conzero} 
\ee 
While the second of these conditions can be seen to hold by a 
simple argument, the first condition depends on details of the coefficients 
$C_m^{(n)}$, which we are currently evaluating (and will be presented 
in \cite{bgk}). 
Assuming that the linear divergence is indeed absent, as it must be, 
we can show that a subleading logarithmic divergence is automatically 
absent since the coefficients also satisfy the sum rule 
\be 
\sum_{n=1}^{8} (-)^{n} (n+3)! C_{n+2}^{(n)} = 0  \: . 
\label{twocon} 
\ee 
 
The remaining integral 
is finite and its value is given by letting $\Lambda \rightarrow \infty$ in 
(\ref{reswilsb}) which gives 
\be 
\langle W (R) \rangle = - 4\pi^2 \int_0^{\infty} dX_T \, X_T^2 \, \sqrt{X_T^2 + R^2} 
\, F(|X_T|). 
\label{remain} 
\ee 
It turns out that this expression reduces to purely boundary terms 
which combine into the  result 
\be 
\langle W (R)\rangle  = -{\pi^{4}\over 2^4}{1\over 4!} 
 \left({g^2_{_{YM}} \over 8\pi^2}\right)^4 e^{2\pi i \tau} 
\sum_{n=2}^{8} \sum_{k=2}^{n} (-)^{n} {(n+3)!\over k+3}  
 C_{n+2}^{(n)} \: , 
\label{mainres} 
\ee 
 where the overall normalization has been reinstated and 
 takes into account the standard $SU(2)$ 
 instanton measure and $\tau$ is the complexified Yang--Mills coupling 
 constant, 
$\tau = \vartheta_{_{YM}}/2\pi + 4\pi i/ 
g^2_{_{YM}}$.  When combined with the complex conjugate contribution due to 
an anti-instanton the result is real but $\vartheta_{_{YM}}$-dependent 
and independent of $R$.

\section{Discussion}

We have seen that the value of the Wilson loop in an instanton 
background can be obtained by direct integration over the supermoduli. 
The calculation made use of the residual $OSp(2,2|4)$ symmetry of the 
configuration.  The expression for the instanton contribution has the 
form of the integral of a divergence over $AdS_5$ 
and therefore the result can be written as a surface term on the 
$\rho=0$ boundary.  More specifically, the support for this integral 
arises from $\rho \sim 0$, $|x_l| \sim R$ and $x_t \sim 0$, which 
means it comes from small instantons touching the loop.  This is just 
the region in which intuitive arguments would suggest the instantons 
could contribute.  Although we have not yet completed the calculation 
 of the exact value of the loop there is no reason to expect it to 
 vanish. Assuming that it is nonzero, 
it would be interesting to identify a topological origin 
for the coefficient characterizing the result. 
The generalization of the semiclassical approximation 
to gauge group $SU(N)$ and arbitrary instanton number in the  large 
$N$ limit is 
straightforward, given the results of \cite{dorey,gk} concerning the ADHM 
construction in this case.

It is far less trivial to make the 
comparison with the type IIB supergravity implied by the 
AdS/CFT correspondence.  The familiar problem is that the supergravity 
limit corresponds to the limit of large 'tHooft coupling, $\lambda = 
g_{_{YM}}^2N >>1$.  The semiclassical 
instanton contributions to certain correlation functions of gauge 
invariant local  operators 
\cite{bgk,dorey} do, in fact, precisely match 
the corresponding D-instanton contributions.  Presumably, this 
agreement means that the leading instanton contributions to these correlation 
functions are independent of the coupling constant.  However, in the case of 
the Wilson loop we know that the perturbative contribution results in 
the behaviour, 
$e^{\lambda} = e^{\sqrt{g_{_{YM}}^2N}}$ and a similar behaviour probably 
also multiplies the instanton contribution.  Clearly, this factor 
reduces to a trivial constant in the semiclassical ($g_{_{YM}} \to 0$) 
limit and is therefore not seen in this approximation. 
 
The presence of a nonzero instanton contribution to the Wilson loop should come 
as no surprise.   After all, the \N4\ theory possesses the highly 
nontrivial Montonen--Olive $SL(2,Z)$ duality symmetry and a 
 separate Wilson loop 
can be defined for each test dyonic particle carrying any of the infinite 
set of electric and magnetic charges that are related by this 
symmetry. Although we have 
not investigated this in detail, it is implausible that these 
relationships can be satisfied without the characteristic dependence on 
$\vartheta_{_{YM}}$ that enters into the instanton terms. 
The precise way in which S-duality is implemented still 
remains to be understood. 
 
As emphasized by Shenker \cite{shen}, matrix models quite generally 
have instantons that are related to the presence of D-instantons in 
string theory.  Although the work of \cite{esz,dg} suggests that there 
is a connection between the Wilson loop and a gaussian matrix model, 
the analysis of \cite{ad} clearly shows that matching the large $N$ 
results does not exclude the presence of a non-trivial 
potential in the matrix model. If this turns out to be the case, 
 the instanton effects discussed in this paper would naturally 
 be identified with those of the matrix model.

\section*{Acknowledgements} 
We are particularly grateful to John Stewart for introducing 
 us to REDUCE, teaching us how to use it and producing the initial 
 programme. We are also grateful to Gary Gibbons, Rajesh Gopakumar,
Nick Manton and  Giancarlo Rossi for useful 
conversations. MB thanks PPARC for support under its visiting 
fellowship scheme.  The work of SK is supported by a Marie Curie 
fellowship.  Partial support of 
a PPARC Special Programme Grant and EEC contract HPRN-2000-00122 
 is also acknowledged.

\end{document}